# NUSTORM FFAG DECAY RING

J.-B. Lagrange*, Imperial College London, UK and FNAL, US
J. Pasternak, Imperial College London and ISIS, RAL, STFC, UK
R. B. Appleby, J. M. Garland, H. Owen, S. Tygier
Cockcroft Institute and Manchester University, UK
A. Bross, A. Liu, FNAL, US

*Abstract*

The neutrino beam produced from muons decaying in a storage ring would be an ideal tool for precise neutrino cross section measurements and search for sterile neutrinos due to its precisely known flavour content and spectrum. In the proposed nuSTORM facility pions would be directly injected into a racetrack storage ring, where circulating muon beam would be captured. The storage ring has two options: a FODO solution with large aperture quadrupoles and a racetrack FFAG (Fixed Field Alternating Gradient) using the recent developments in FFAGs. Machine parameters, linear optics design and beam dynamics are discussed in this paper.

## INTRODUCTION

The production of a neutrino beam with a defined spectrum and flux composition using a muon decay is a well-established idea. This concept was developed into the Neutrino Factory facility proposal, which was then addressed in several dedicated research and development studies culminating in the International Design Study for the Neutrino Factory (IDS-NF) [1]. The neutrinos from STORed Muon beam (nuSTORM) project [2] is the simplest implementation of a neutrino factory, with high energy pions produced at the target directly injected into the ring after passing through a short transfer line equipped with a chicane to select charge of the beam. Once in the ring, decaying pions will form the muon beam. A fraction of the muon beam with energy lower than the injected parent pions will be stored in the ring and a fraction with similar or larger energy will be extracted with a mirror system of the injection at the end of the long straight section to avoid activation in the arc. They may also be used for accelerator research and development studies for future muon accelerators.

As the flux intensity is one of key elements for a successful neutrino experiment, it is proposed to push the momentum acceptance of the ring to ±8% or even ±19%. Two design are developed in parallel, the first one based on the standard accelerator lattice with separated function magnets [3] and the designs based on scaling FFAG lattice. The scaling FFAG technology allows to have zero chromaticity with large dynamical acceptance, which enables large momentum spread of the beam with low losses by avoiding the dangerous resonances. This paper describes the racetrack FFAG (RFFAG) ring design for nuSTORM.

* j.lagrange@imperial.ac.uk

## RING DESIGN

### Quadruplet Cell in the Straight Section

The RFFAG ring design consists of long straight sections pointing towards neutrino detectors, where the majority of the pions will decay into muons and along which the neutrino beam will be formed firstly from the pion decay and secondly from the muon decay. Both signals can be separated by the detector timing information. The ring contains also compact arcs in order to achieve a large neutrino beam production efficiency minimizing the size of the ring and the associated cost. An RFFAG design based on triplet cell with long drift spaces was proposed previously [4]. However a quadruplet cell DFFD gives an additional long straight between the two F magnets, increasing the efficiency of the neutrino production towards the detector. Furthermore, since D and F magnets are identical, only one type of magnet needs to be designed and manufactured.

### Lattice Design

The ring consists of several distinct cell types:

- Straight scaling FFAG cells in the neutrino production straight sections with room temperature magnets, in which the vertical magnetic field on the median plane follows an exponential law [5].
- Regular scaling FFAG arc cells equipped with the superconducting combined function FFAG-type magnets, in which the vertical magnetic field on the median plane follows the circular scaling FFAG law [6]. The horizontal phase advance is chosen so that the arc is transparent, which means Pi phase advance, which helps to minimize beta beating, while FD ratio is chosen to adjust the vertical beta function.
- The matching arc cells are based on superconducting FFAG, with transparent horizontal phase advance, while the radius of the cells are chosen carefully to adjust the dispersion between the straight section and the arc part. At this place, 1.34 meter dispersion gives sufficient beam separation between circulating maximum muon momentum (4.4 GeV/c) and injected minimum pion momentum (4.5 GeV/c). A 2.6 meter-long drift space is then allowed to put a septum magnet for injection.

### Tune Scan for Dynamical Acceptance Study

To study the quadruplet solution with respect to the dynamical acceptance, an automated procedure to change the tune point of the lattice has been used. Stepwise tracking using Runge Kutta integration in field model with Enge-type fringe fields has been used for this study. The results are presented in Fig. 1. The choice of the tune point was done through two parameters, the dynamical acceptance in both horizontal and vertical and the muon capture efficiency in the straight section. The muon capture efficiency is a function of the dispersion in the straight cells, giving the distance between the reference orbit of injected pion beam and and the reference orbit of the muon beam, and the maximum stable horizontal amplitude the lattice can accept. The chosen lattice in this scan gives a horizontal emittance of about 2 mm, a vertical emittance of about 1 mm, and a maximum stable horizontal amplitude larger than the distance between the reference orbit of injected pion beam and and the reference orbit of the muon beam.

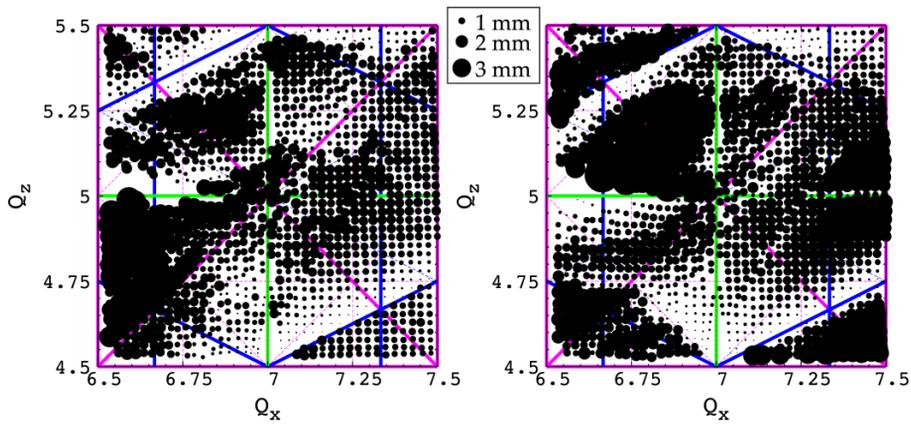

Figure 1: Results of a scan showing the horizontal (left) and vertical (right) acceptances for various positions in the tune diagram. The area of each circle is proportional to the transverse (either horizontal or vertical) acceptance. Normal resonances lines, plotted up to the octupole with structure resonances in bold, are superimposed.

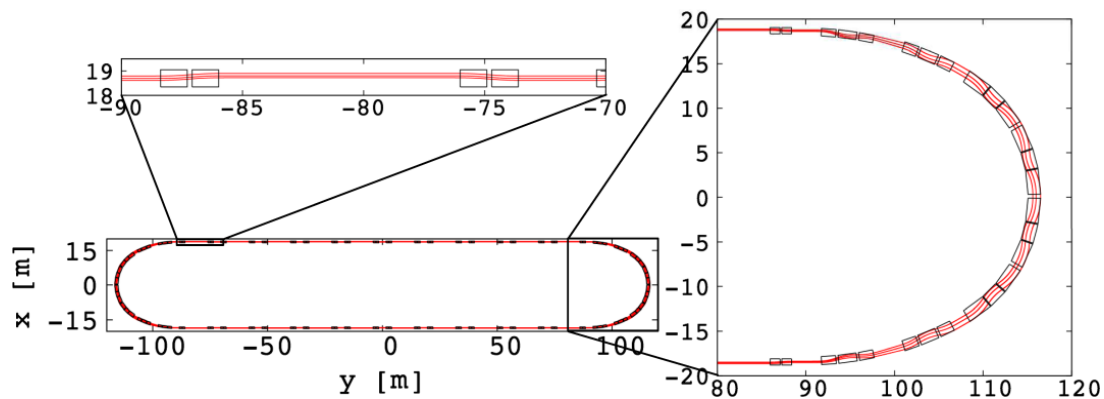

Figure 2: Top view of the racetrack FFAG lattice (bottom left scheme). The top left scheme shows a zoom on the straight section and the right scheme a zoom on the arc section. Matched, minimum and maximum momenta muon closed orbits are shown in red. Effective field boundaries with collimators are shown in black.

*Optimized Lattice*

Ring parameters are summarized in Table 1. Closed orbits of matching momentum, minimum momentum and maximum momentum are shown in Fig. 2. Dispersion and beta-functions at matching momentum are shown in Fig. 3. Magnetic field for maximum momentum muon closed orbit is presented in Fig. 4. Stability of the ring tune has been studied over ±19% momentum range. The tune shift is presented in Fig. 5.

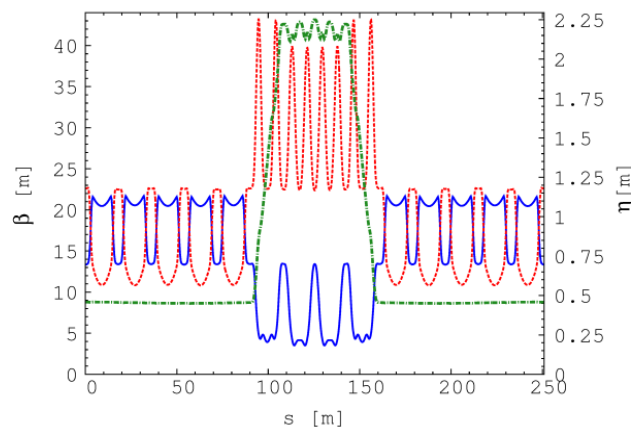

Figure 3: Horizontal (plain blue), vertical (dotted red) periodic betatron functions (left scale) and dispersion (mixed green line, right scale) of half of the ring for matching momentum. The plot is centered on the arc part.

Table 1: Lattice Parameters

| Parameter | Value |
|---|---|
| Total circumference | 502 m |
| Length of one straight section | 180 m |
| One straight section/circumference ratio | 36% |
| Momentum acceptance | 3.7 GeV/c ±19% |
| Ring tune (H, V) | (7.18, 4.88) |
| Number of cells in the ring: | |
| Straight cells | 20 |
| Arc matching cells | 8 |
| Regular arc cells | 8 |
| m-value in straight cells | 2.2 m$^{-1}$ |
| Packing factor in straight cells | 0.24 |
| Max. scallop angle in straight cells | 76 mrad |
| k-value in regular arc cells | 6.056 |
| $R_0$ in regular arc cells | 16.4 m |
| Packing factor in regular arc cells | 0.92 |
| k-value in matching cells | 26.0 |
| $R_0$ in matching cells | 36.15 m |
| Packing factor in matching cells | 0.57 |

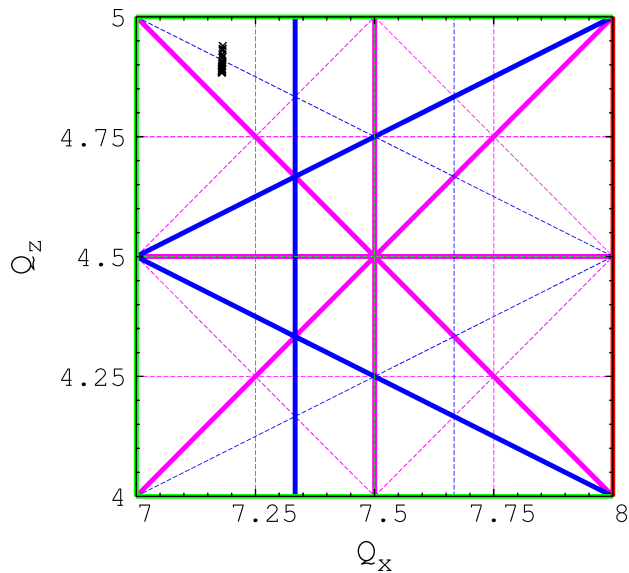

Figure 5: Tune diagram for momenta ±19% around 3.7 GeV/c. Integer (red), half-integer (green), third integer (blue) and fourth integer (purple) normal resonances are plotted. Structural resonances are in bold.

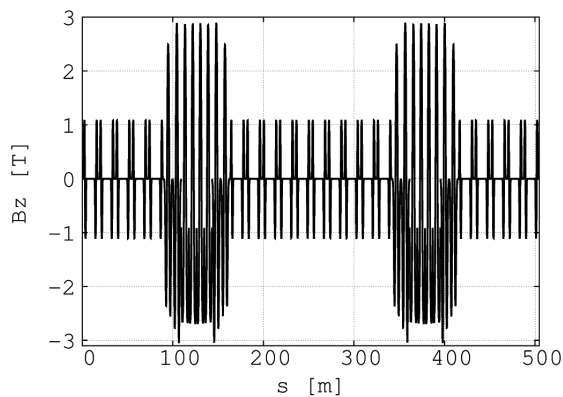

Figure 4: Vertical magnetic field on the median plane for the maximum circulating muon momentum of half of the ring. The plot is centered on the arc part.

## SUMMARY

nuSTORM project allows to address essential questions in the neutrino physics, in particular by offering the best possible way to measure precisely neutrino cross sections and by allowing to search for light sterile neutrinos. It would also serve as a proof of principle for the Neutrino Factory and can contribute to the R&D for future muon accelerators. The FFAG solution for the decay ring gives good performance regarding transverse and momentum acceptance. A neutrino flux computation with this solution remains to be done. A full comparison between separated function lattice and RFFAG regarding neutrino flux performance is also planned.

## ACKNOWLEDGMENT

This work was supported by the US Department of Energy (DOE) and the UK Science and Technology Facilities Council (STFC) in the PASI (Proton Accelerators for Science and Innovation) framework, including the grants ST/G008248/1 and ST/K002503/1. Authors wish to acknowledge this support.

## REFERENCES

[1] IDS-NF website: http://ids-nf.org/

[2] D. Adey et al., "nuSTORM - Neutrinos from STORed Muons: Proposal to the Fermilab PAC", arXiv:1308.6822.

[3] A. Liu et al., "Decay Ring Design Update for nuSTORM", in *Proc. of IPAC'14*, Dresden, Germany, paper TUPRI006, 2014.

[4] J-B. Lagrange et al., "Progress on the design of the racetrack FFAG decay ring for nuSTORM", in *Proc. of IPAC'15*, Richmond, Virginia, USA, paper WEPWA043, 2015.

[5] J.-B. Lagrange et al., "Straight scaling FFAG beam line", in *Nucl. Instr. Meth. A*, vol. 691, pp. 55–63, 2012.

[6] K. R. Symon, D. W. Kerst, L. W. Jones, L. J. Laslett, K. M. Terwilliger, Fixed-Field Alternating-Gradient Particle Accelerators, in *Phys. Rev.*, vol. 103, no. 6, pp. 1837–1859, 1956.